\documentstyle[aps,prl,epsf,floats,twocolumn]{revtex}
\begin{document}
\draft
\twocolumn[\hsize\textwidth\columnwidth\hsize\csname @twocolumnfalse\endcsname
\title{Effect of Local Magnetic Moments on the Metallic Behavior in Two 
Dimensions}
\author{X. G. Feng$^1$, Dragana Popovi\'{c}$^{1}$, and S. Washburn$^{2}$}
\address{$^{1}$National High Magnetic Field Laboratory, 
Florida State University, Tallahassee, FL 32310 \\ 
$^{2}$Dept. of Physics and 
Astronomy, The University of North Carolina at Chapel Hill, Chapel Hill, NC 
27599 }
\date{\today}
\maketitle

\begin{abstract}

The temperature dependence of conductivity $\sigma (T)$ in the metallic phase 
of a two-dimensional electron system in silicon has been studied for different
concentrations of local magnetic moments.  The local moments have been induced
by disorder, and their number was varied using substrate bias.  The data
suggest that in the limit of $T\rightarrow 0$ the metallic behavior, as
characterized by $d\sigma/dT < 0$, is suppressed by an arbitrarily small 
amount of scattering by local magnetic moments.

\end{abstract}

%
%
%
%
]

A metal-insulator transition (MIT) has been observed recently in a variety of
two-dimensional (2D) electron~\cite{Krav,DP_MIT,AlAs_elect,GaAs_elect} and 
hole~\cite{SiGe_holes,GaAs_holes} systems but there is still no generally 
accepted microscopic description of the 2D metallic phase.  Some of the 
relevant properties of the 2D metal include: (a) an increase of conductivity 
$\sigma$ with decreasing temperature $T$ (i.~e. $d\sigma/dT < 0$) for carrier 
densities $n_s > n_c$ ($n_c$ -- critical density); and (b) a suppression of 
the $d\sigma/dT < 0$ behavior by magnetic field~\cite{Bfield}.  The latter 
suggests the importance of the spin degrees of freedom, which can be probed 
further by studying the effect of local magnetic moments on the transport 
properties of the conduction electrons.  Indeed, magnetic impurities have been
used extensively over the last several decades to probe the properties of 
metals, and continue to be relevant today in attempts to understand 
heavy-fermion materials and high-$T_c$ superconductors~\cite{Hewson}.  In the 
experiment discussed below, the number of local moments is varied in a 
controlled way.  We show that in the $T\rightarrow 0$ limit the $d\sigma/dT 
< 0$ behavior is suppressed by an arbitrarily small amount of scattering of 
the conduction electrons by disorder-induced local moments.  

We present results obtained on a 2D electron system in Si 
metal-oxide-semiconductor field-effect transistors (MOSFETs).  In such a
device,  the disorder is due to the oxide charge scattering (scattering by 
ionized impurities randomly distributed in the oxide within a few \AA\, of the
interface) and to the roughness of the Si-SiO$_2$ interface~\cite{AFS}.  For a
fixed $n_s$, it is possible to
change the disorder by applying the substrate (back gate) bias $V_{sub}$.  In
particular, the reverse (negative) $V_{sub}$ moves the electrons closer to the
interface, which increases the disorder.  It also increases the splitting
between the subbands since the width of the triangular potential well at the
interface is reduced by applying negative $V_{sub}$.  Usually, only the lowest
subband is occupied at low $T$, giving rise to the 2D behavior.  In 
sufficiently disordered samples, however, the band tails associated with the
upper subbands~\cite{tails} can be so long that some of their strongly 
localized states may be populated even at low $n_s$, and act as additional
scattering centers for 2D electrons.  Clearly, the negative $V_{sub}$ reduces
this type of scattering by depopulating the upper subband.  The effect of 
scattering by electrons localized deep
in the tails of the upper subband was first observed as an enhancement of the 
mobility $\mu$ at low $n_s$ with negative $V_{sub}$~\cite{Alan}, and was 
subsequently studied in more detail by other groups using different 
measurements and techniques~\cite{subbands}.  More recently, we have used one
negative value of $V_{sub}$ to enhance $\mu$ (reduce the disorder) at 
intermediate values of $n_s$, and observed the change from $d\sigma/dT >0$ to
$d\sigma/dT < 0$ for $n_s > n_c$~\cite{DP_MIT}.  Here, however, we present a 
systematic study of this process as the disorder is varied using $V_{sub}$.  We
show clearly that the bare value of high $T$ (Drude) mobility is {\em not}
sufficient to predict the sign and the magnitude of $d\sigma/dT$ at low $T$
but rather that it is the {\em type} of the disorder that is 
relevant~\cite{Shayegan}.  In particular, we show that scattering by electrons
localized in the tail of the upper subband has a much more profound effect on 
$d\sigma/dT$ than potential scattering due to oxide charges and surface 
roughness.  This is attributed to spin flip scattering by electrons in 
localized states that are singly populated due to a strong on-site Coulomb 
repulsion, and act as local magnetic moments.  Large on-site Coulomb 
interaction has been well documented in systems similar to ours, such as 
electrons in quantum dots~\cite{dots}, and other materials with strongly 
localized states~\cite{Beasley}.  For typical localization lengths of 
$\sim$100~\AA\, in Si MOSFETs~\cite{AFS,Timp}, the on-site Coulomb repulsion is
$\sim$10~meV.  Therefore, such states will be singly occupied at low 
$n_s$~\cite{AFS}.

Our measurements were carried out on n-channel Si MOSFETs with the oxide 
charge density of $3\times 10^{10}$cm$^{-2}$, determined using standard 
techniques~\cite{AFS,thresholds}.  Other details of the sample structure are
given in Ref.~\cite{DP_MIT}.  For a fixed $V_{sub}$, $n_s$ was controlled by
the gate voltage $V_g$ and determined in a standard 
fashion~\cite{AFS,thresholds}.  $\sigma (V_{g})$ was measured at temperatures
$0.3 < T < 4.5$~K for $n_s$ of up to $3\times 10^{12}$cm$^{-2}$ and for
--50~V$\leq V_{sub}\leq$+1~V.  The effect of $V_{sub}$ on $\mu$ at 4.2~K
was found~\cite{icps24} to be consistent with earlier work and our 
interpretation.   In particular, for $n_s < n_{max}$ ($n_{max}\sim 5\times 
10^{11}$cm$^{-2}$ is the density where $\mu$ reaches its maximum), an increase
of $\mu$ is observed~\cite{icps24} with the negative $V_{sub}$ as a result of 
the decreased scattering by local moments from the upper subband, and 
consistent with early work~\cite{Alan}.  For $n_s > n_{max}$, $\mu$ decreases 
with (negative) $V_{sub}$, consistent with the fact that surface roughness 
scattering is the dominant source of disorder in this range of 
$n_s$~\cite{AFS}.  This is a result of an increased proximity of the 2D 
electrons to the interface and, possibly, a reduction in scattering by local
moments from the upper subband.  The latter could be due to a smaller number of
local moments being present in a sample at high $n_s$ (for a given $V_{sub}$) 
because of an improved screening by 2D electrons.  For sufficiently high 
negative $V_{sub}$ ($-V_{sub} > 35$~V), the 4.2~K mobility decreases with 
$V_{sub}$ for all $n_s$, suggesting that the upper subband has been completely
depopulated and that the further increase in $V_{sub}$ leads only to 
increasing disorder due to potential scattering from roughness at the 
Si-SiO$_2$ interface.

\begin{figure}
\epsfxsize=3.2in \epsfbox{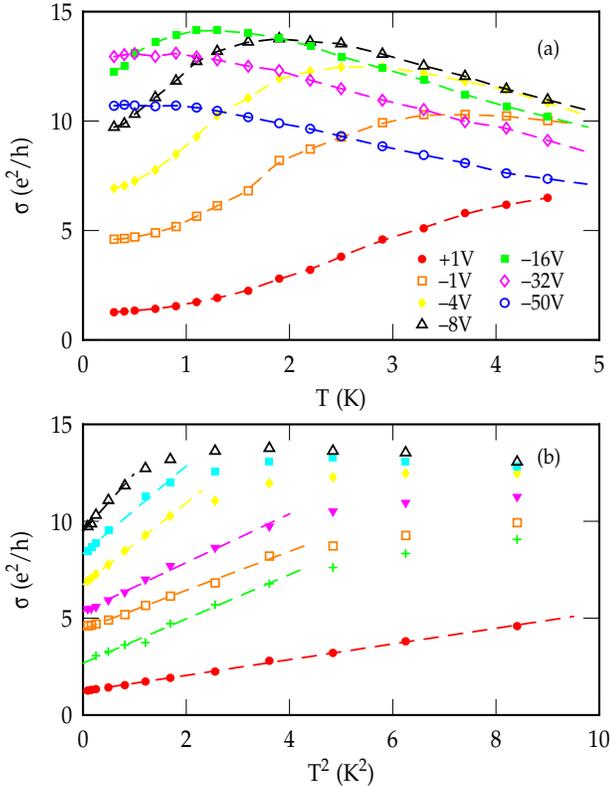}\vspace{5pt}
\caption{Temperature dependence of the conductivity $\sigma$ for 
$n_{s}=3.0\times 10^{11}$cm$^{-2}$.  (a) The data are shown for different 
values of $V_{sub}$ as given on the plot.  (b) The data are plotted vs. $T^2$,
and shown for $V_{sub}=+1,-0.5,-1,-2,-4,-6,-8$~V going from bottom to
top.
\label{sigma}}
\end{figure}

Fig.~\ref{sigma}(a) shows some typical results for $\sigma (T)$ in the $n_s <
n_{max}$ range as a function of $V_{sub}$.  The metallic behavior, such
that $d\sigma/dT < 0$, spreads out towards lower $T$ with the increasing 
negative $V_{sub}$, i.~e. as the scattering by local moments is reduced, and it
also spreads out towards higher values of $n_s$ (not shown).  In other words, 
$\sigma (T)$ displays a maximum at $T=T_m$, such that $T_m$ shifts to lower 
$T$ with the (negative) $V_{sub}$.  As $(-V_{sub})$ is increased beyond 35~V, 
the form of $\sigma (T)$ is no longer very sensitive to changes in $V_{sub}$ 
even though the disorder due to potential scattering increases.  In addition, 
by comparing the data for $V_{sub}=-50$~V and --1~V, for example, it is 
obvious that $d\sigma/dT < 0$ behavior is more pronounced ($T_m$ is lower) 
when scattering by local moments is reduced {\em even though the 4.2~K 
mobility is lower}~\cite{Shayegan}.  This demonstrates clearly the need to 
distinguish between different types of disorder, a fact that has been 
overlooked in some theoretically proposed phase 
diagrams~\cite{theories11,theories8}.  For $T<T_m$, $\sigma$ decreases with 
decreasing $T$ and, in fact, follows a $T^2$ form at the lowest $T$ 
[Fig.~\ref{sigma}(b)].  Such $\sigma (T)$ is often considered to be a 
signature of local magnetic moments, and results from the Kondo 
effect~\cite{Hewson}.  Here it represents a direct evidence for the existence 
of local moments in our samples.  A detailed study of this regime has been 
presented elsewhere~\cite{ourKondo}.  Fig.~\ref{sigma}(b) also shows that the 
range of $T$ ($T<T_m$) where local moments dominate transport becomes smaller 
as their number is reduced by increasing negative $V_{sub}$.

\begin{figure}
\epsfxsize=3.2in \epsfbox{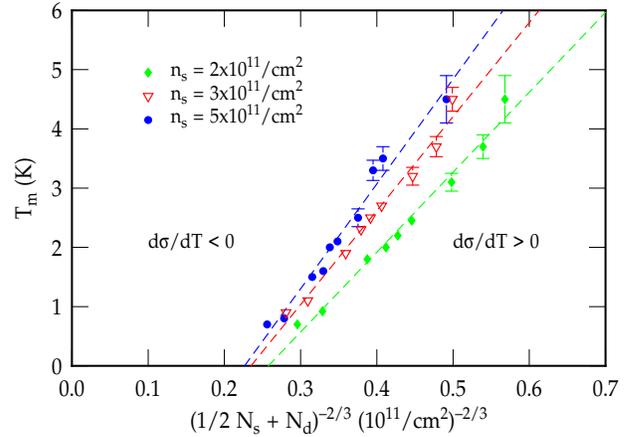}\vspace{5pt}
\caption{Position of the maximum $T_m$ in $\sigma (T)$ for different values of
$n_s$ as a function of the inverse subband splitting.  (The left hand side on 
the x-axis corresponds to large values of the negative $V_{sub}$, and the 
right hand side corresponds to low values of $V_{sub}$.) The metallic behavior
with $d\sigma/dT < 0$ is observable at $T>T_m$.  $T_m$ extrapolates to zero 
(dashed lines) for $V_{sub}\sim -40$~V, corresponding to the subband splitting
of the order of 30~meV.
\label{tm}}
\end{figure}

The position of the maximum $T_m$ in $\sigma (T)$ is shown in 
Fig.~\ref{tm} for different values of $n_s$ as a function of the inverse 
subband
splitting ($N_d$ is the depletion layer charge density, which increases with 
the reverse $V_{sub}$)~\cite{AFS}.  The metallic behavior with $d\sigma/dT < 
0$ is observable at $T > T_m$ [see Fig.~\ref{sigma}(a)].  For a given $n_s$, 
$T_m$ extrapolates to zero for a finite value of the subband splitting.  This 
value 
is slightly higher for higher $n_s$, consistent with the fact that the number 
of local moments in the upper subband is also slightly higher (Fermi energy 
$E_F$ is higher) and, therefore, one needs to apply more $V_{sub}$ in order to 
depopulate the upper subband.  From the data shown in Fig.~\ref{tm}, it
follows that $T_m$ goes to zero for $V_{sub}\sim -40$~V in agreement with the
measurement of the 4.2~K mobility as discussed above.  $V_{sub}\sim -40$~V
corresponds to the subband splitting of the order of 30~meV~\cite{AFS}.  The 
extent of the tail of the upper subband derived in this way is consistent with
earlier work~\cite{subbands}.  Our data, therefore, show that the 2D metal 
with $d\sigma/dT < 0$ can exist at $T=0$ only in the absence of scattering by 
disorder-induced local moments.  This is similar to the behavior observed in a
magnetic field~\cite{Bfield}, and consistent with some 
theoretical models~\cite{theories4,Finkelstein,CCLM,theories5}.

\begin{figure}[t]
\epsfxsize=3.4in \epsfbox{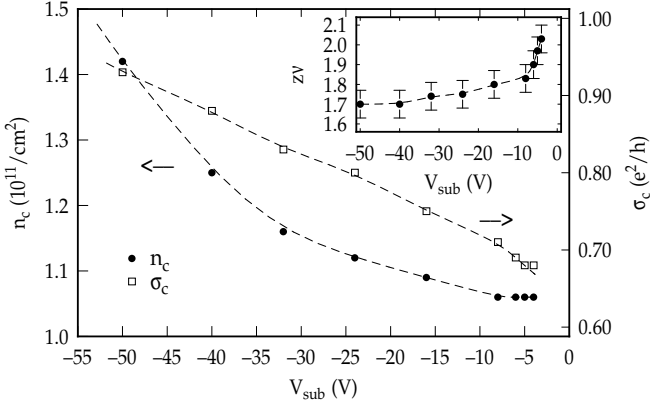}\vspace{5pt}
\caption{Critical density $n_c$ and critical conductivity $\sigma_c$ determined
from the data at $T > T_m$ as a function of $V_{sub}$.  The disorder due to 
potential scattering increases with the negative $V_{sub}$.  Inset: critical 
exponents $z\nu$ vs. $V_{sub}$ obtained from the same data.  Dashed lines guide
the eye.
\label{critical}}
\end{figure}

In the presence of scattering by local moments, $d\sigma/dT < 0$ is observable
at $T > T_m$, where $T_m$, of course, can be arbitrarily small.  For a fixed 
$V_{sub}$, $\sigma (n_s,T)$ for $T > T_m$ exhibits~\cite{DP_MIT} all of the 
properties of a 2D MIT.  Fig.~\ref{critical} shows the values of $n_c$ and the
(apparent)~\cite{sigmac}
critical conductivity $\sigma_c$ determined only from the data at $T > T_m$,
where scattering by local moments is not significant, as a function of 
$V_{sub}$.  
Both $n_c$ and $\sigma_c$ increase monotonically with the reverse $V_{sub}$, 
i.~e. with an increase in disorder due to potential scattering.  The increase 
of $n_c$ and $\sigma_c$ with disorder observed here on a single sample is in 
agreement with the same conclusion reached by comparing MOSFETs with different
peak mobilities~\cite{Pudalov}.  Fig.~\ref{critical} inset shows the 
dependence of the critical exponents $z\nu$  ($z$-- dynamic exponent, $\nu$-- 
correlation length exponent) on $V_{sub}$.  $z\nu$ was determined by scaling 
the data in the vicinity of $n_c$, where $\sigma(n_s,T)=\sigma_c
f(T/\delta_{n}^{z\nu})$ [$\delta_n=(n_s-n_c/n_c)$]~\cite{DP_MIT,sigmac}.

\begin{figure}[t]
\epsfxsize=3.3in \epsfbox{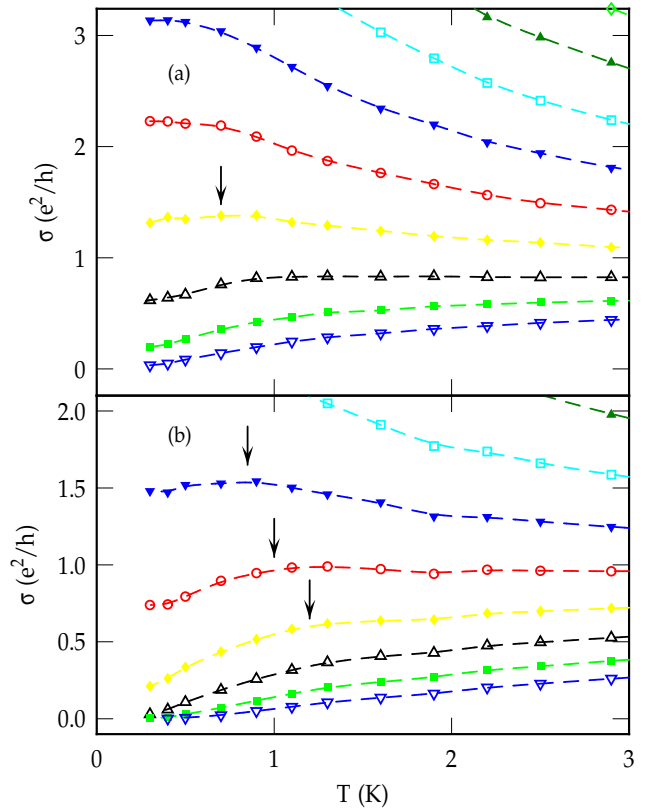}\vspace{5pt}
\caption{Temperature dependence of conductivity for $n_s$ from $1.02\times
10^{11}$cm$^{-2}$ (bottom curve) up, in steps of $0.1\times 10^{11}$cm$^{-2}$ 
and (a) $V_{sub}=-40$~V, (b) $V_{sub}=-50$~V.  The arrows indicate the peaks in
$\sigma (T)$ for $n_s (10^{11}$cm$^{-2})=1.42$ and 1.52 for $V_{sub}=-50$~V, 
and possible peaks for $n_s=1.32\times 10^{11}$cm$^{-2}$ at both $V_{sub}$.
\label{data}}
\end{figure}

Fig.~\ref{data}(a) shows $\sigma (T)$ for $V_{sub}=-40$~V for a small range of
$n_s$ close to $n_c$.  Based on the analysis discussed above, we
are confident that for this value of $V_{sub}$ there are no local moments 
associated with the upper subband in the sample.  Indeed, the $\sigma (T)$ 
curves do not exhibit a maximum in the measured range of $T$, except possibly
where shown in Fig.~\ref{data}(a).  Such a structure might, in fact, be due to
local moments associated with the lowest subband.  From the theoretical point 
of view, the possibility of disorder-induced local moment formation in 
strongly interacting electronic systems has been suggested since the early 
developments of the theory of interacting disordered 
systems~\cite{oldmoments,Finkelstein} and has been studied, subsequently, 
using different models~\cite{newmoments}.  In order to test this idea, we have
increased $V_{sub}$ to $-50$~V, thus increasing the disorder, which is now due
only to scattering by the oxide charges and the surface roughness.  
Fig.~\ref{data}(b) shows that, for the same $n_s$, the values of $\sigma$ are 
reduced as expected for higher disorder.  More importantly, we observe the 
{\em appearance of the peak} in $\sigma (T)$ for those $n_s$ for which it was
clearly absent in Fig.~\ref{data}(a).  Also, a weak structure (peak) in 
$\sigma (T)$ for $n_s=1.32\times 10^{11}$cm$^{-2}$ has shifted to higher $T$. 
This increase of $T_m$, and the dependence of $T_m$ on $n_s$ as shown in 
Fig.~\ref{data}(b), are qualitatively the same as what was observed by 
increasing the number of local moments associated with the upper subband.  
These results strongly suggest that an increase in the potential 
(non-magnetic) scattering has led to the formation of the (additional) local 
moments in the system.  We have also observed similar peaks in $\sigma (T)$ in
different Si MOSFETs at $T < 0.4$~K but, without a systematic study such as 
this one, it would have been impossible to determine their origin.  We 
speculate that local moments might exist in other materials as well but that 
the corresponding $T_m$ might be experimentally inaccessible in high-mobility 
devices.

Back gate bias was used recently in a 2D hole system~\cite{Shayegan_rashba} to
study the effect of the spin-splitting due to the spin-orbit interaction and
the inversion asymmetry of the confining potential~\cite{Rashba}.  It was found
that the magnitude of the $d\sigma/dT < 0$ behavior was reduced as the 
spin-splitting decreased, i.~e. as the confining potential became more 
symmetric.  In our samples, we observe the opposite: the triangular confining 
potential becomes more symmetric with the application of the reverse 
$V_{sub}$, and that is exactly when the $d\sigma/dT < 0$ behavior appears.  
Therefore, even if the effect of the spin-orbit interaction exists in our 
samples, it does not drive the MIT.

The reverse $V_{sub}$ also reduces the average spatial extent $\Delta z$ of the
inversion layer charge density in the direction perpendicular to the 
interface (typically, $\Delta z\approx 20-30$~\AA)~\cite{AFS}, leading to an
increase of the effective Coulomb interaction~\cite{Sarma}.  Since 
$r_s$ ($r_s$ -- the average inter-carrier separation in units of the effective
Bohr radius) is already fairly large ($r_s\sim 15$) for $n_s$ close to $n_c$
(given in Fig.~\ref{critical}), we expect that the further increase in the 
Coulomb interaction for a fixed $n_s$ would only lead to an insulating 
behavior (see, e.~g. Refs.~\cite{theories11,theories8}) and not to the metallic
behavior with $d\sigma/dT < 0$, as observed.  We conclude that the effect of
$V_{sub}$ on the effective Coulomb interaction is not the dominant effect in 
our samples.

Our study shows that the 2D metal with $d\sigma/dT < 0$ can exist in the 
$T\rightarrow 0$ limit only in the absence of scattering by local magnetic 
moments.  Our results emphasize the key role of the spin degrees of freedom in
the physics of the low density 2D electron system.

The authors are grateful to V. Dobrosavljevi\'{c} and A. B. Fowler for helpful 
discussions.  This work was supported by NSF Grant No. DMR-9796339.


\begin{references}
\bibitem{tails} Usually, this is a tail of the $E_{0'}$ subband -- the lowest
subband associated with the four conduction band valleys with the light 
effective mass (0.190~$m_e$) perpendicular to the interface.  See also 
Ref.~\cite{AFS}.
\bibitem{Shayegan} This also explains the recent observation 
(Refs.~\cite{icps24} and \cite{Shayegan_rashba}) that the metallic behavior in
some samples is more pronounced (i.~e. the magnitude of negative $d\sigma/dT$ 
is larger) when the mobility is lower.
\end{references}
\end{document}